\documentclass[12pt]{article}
\usepackage{a4,graphicx}
\usepackage[small,bf]{caption}

\def\lsim{\mathrel{\rlap{\lower4pt\hbox{\hskip1pt$\sim$}}
    \raise1pt\hbox{$<$}}}                
\def\gsim{\mathrel{\rlap{\lower4pt\hbox{\hskip1pt$\sim$}}
    \raise1pt\hbox{$>$}}}                

\begin{document}

\title{
\vspace{-3.25cm}
\flushright{\small ADP-15-25/T927} \\
\vspace{-0.35cm}
{\small DESY 15-154} \\
\vspace{-0.35cm}
{\small Edinburgh 2015/19} \\
\vspace{-0.35cm}
{\small Liverpool LTH 1054} \\
\vspace{-0.35cm}
{\small August 24, 2015}  \\
\vspace{0.5cm}
\centering{\Large \bf Wilson flow and scale setting from lattice QCD}}

\author{\large
         V.~G. Bornyakov$^a$, R. Horsley$^b$, 
         R. Hudspith$^c$, Y. Nakamura$^d$,        \\
         H. Perlt$^e$, D. Pleiter$^f$, 
         P.~E.~L. Rakow$^g$, G. Schierholz$^h$,   \\
         A. Schiller$^e$, H. St\"uben$^i$ and J.~M. Zanotti$^j$ \\[1em]
         -- QCDSF-UKQCD Collaboration -- \\[1em]
        \small $^a$ Institute for High Energy Physics, 
               Protvino, \\[-0.5em]
        \small 142281 Protvino, Russia, \\[-0.5em]
        \small Institute of Theoretical and Experimental Physics,
               Moscow, \\[-0.5em]
        \small 117259 Moscow, Russia, \\[-0.5em]
        \small School of Biomedicine, 
               Far Eastern Federal University, \\[-0.5em]
        \small 690950 Vladivostok, Russia      \\[0.25em]
        \small $^b$ School of Physics and Astronomy,
               University of Edinburgh, \\[-0.5em]
        \small Edinburgh EH9 3FD, UK \\[0.25em]
        \small $^c$ Department of Mathematics and Statistics,
               York University,  \\[-0.5em]
        \small Toronto, ON Canada M3J 1P3 \\[0.25em]
        \small $^d$ RIKEN Advanced Institute for
               Computational Science, \\[-0.5em]
        \small Kobe, Hyogo 650-0047, Japan \\[0.25em]
        \small $^e$ Institut f\"ur Theoretische Physik,
               Universit\"at Leipzig, \\[-0.5em]
        \small 04109 Leipzig, Germany \\[0.25em]
        \small $^f$ J\"ulich Supercomputing Centre,
               Forschungszentrum J\"ulich, \\[-0.5em]
        \small 52425 J\"ulich, Germany, \\[-0.5em]
        \small Institut f\"ur Theoretische Physik,
               Universit\"at Regensburg, \\[-0.5em]
        \small 93040 Regensburg, Germany \\[0.25em]
        \small $^g$ Theoretical Physics Division,
               Department of Mathematical Sciences, \\[-0.5em]
        \small University of Liverpool,
               Liverpool L69 3BX, UK \\[0.25em]
        \small $^h$ Deutsches Elektronen-Synchrotron DESY, \\[-0.5em]
        \small 22603 Hamburg, Germany \\[0.25em]
        \small $^i$ Regionales Rechenzentrum, Universit\"at Hamburg, \\[-0.5em]
        \small 20146 Hamburg, Germany \\[0.25em]
        \small $^j$ CSSM, Department of Physics,
               University of Adelaide, \\[-0.5em]
        \small Adelaide SA 5005, Australia}

\date{}

\maketitle


\clearpage

\begin{abstract}
We give a determination of the phenomenological value of the 
Wilson (or gradient) flow scales $t_0$ and $w_0$ for $2+1$
flavours of dynamical quarks. The simulations are performed
keeping the average quark mass constant, which allows the approach
to the physical point to be made in a controlled manner.
$O(a)$ improved clover fermions are used and together with
four lattice spacings this allows the continuum extrapolation to be taken.
\end{abstract}







\section{Introduction} 


Numerical lattice QCD simulations naturally determine dimensionless
quantities such as mass ratios and matrix element ratios,
however determining a physical value requires the introduction of a scale,
usually taken from experiment. A hadron mass, such as the proton mass,
or decay constant, such as the pion decay constant,
are often used for this purpose. We discuss here setting the scale using
flavour-singlet quantities, which in conjunction with simulations
keeping the average quark mass constant allow $SU(3)$ flavour breaking
expansions to be used. This is illustrated here using $2+1$ clover fermions,
and a determination of the Wilson flow scales $t_0$ and $w_0$ is given.
These are `secondary' scales and are not experimentally accessible and
thus they have to be matched to physical quantities. These flow scales
are cheap to compute from lattice simulations (for example they do not
require a knowledge of quark propagators) and accurate (for example
they do not require a determination of the potential which requires
the limit of a large distance). So once the phenomenological value
of the flow scales is known the determination of physical values
becomes more tractable.

Flow and flow variables were introduced by L\"uscher, \cite{luscher09a}.
We follow him here, \cite{luscher10a}, in particular in our brief discussion
of the $t_0$ scale. Flow represents a smoothing of the gauge fields.
We denote the flow time by $t$, and the link variables at this time
by $U_\mu(x,t)= \exp(iT^a\theta_\mu^a(x,t))$ which evolve according to
\begin{eqnarray}
   {dU_\mu(x,t) \over dt} 
      = i T^a \, {\delta S_{\rm flow}[U] \over \delta \theta_\mu^a(x,t)} \,
                                 U_\mu(x,t)\,, 
                       \quad \mbox{with}\,\,\, U_\mu(x,0) = U_\mu(x) \,,
\end{eqnarray}
with $S_{\rm flow}[U]$ being the flow action, which does not have to be the
same as the action used to generate the gauge variable. ($x$ is just
the normal $4$-dimensional Euclidean space-time.) Setting 
\begin{eqnarray}
   F(t) \equiv t^2 \langle E(t) \rangle  \,,
        \quad \mbox{where} \,\,\, E(t) 
       = \textstyle{1 \over 4} F_{\mu\nu}^{a\,2}(t) \,,
\end{eqnarray}
then we define the $t_0$ scale by
\begin{eqnarray}
   \left. F(t) \right|_{t=t_0(c)} = c \,.
\end{eqnarray}
Alternatively, \cite{borsanyi12a} define the $w_0$ scale as
\begin{eqnarray}
   \left. t {d \over dt} F(t) \right|_{t=w_0^2(c)} = c \,,
\end{eqnarray}
where in both definitions $c$ is a constant, conventionally
taken as $c = 0.3$. We require a value of $c$ such that
$a \ll \sqrt{8t_0} \ll L$, $L$ being the lattice size and
this value was found to be a suitable choice, \cite{luscher10a}.
Alternative suggestions have been made, see e.g.\  \cite{sommer13a,asakawa15a}.

In this article we shall give a determination of $\sqrt{t_0^{\exp}}$ and
$w_0^{\exp}$. There have been several determinations for different numbers
of flavours. These include quenched $n_f = 0$ or quenched,
\cite{asakawa15a,francis15a}; $n_f = 2$, \cite{datta15a,bruno13a};
$n_f = 2+1$, \cite{bazavov14a,borsanyi12a,blum14a}; $n_f = 2+1+1$, 
\cite{bazavov15a,dowdall13a,deuzeman12a}.
We have also published preliminary results, \cite{horsley13a}.

The plan of this article is as follows. In section~\ref{extrap_singlet}
we describe our method of approaching the physical quark mass starting
from a point on the $SU(3)$ flavour symmetric line, 
\cite{bietenholz10a,bietenholz11a}. We also discuss the general property
of singlet quantities, that they have a stationary point about this
$SU(3)$ flavour symmetric line.
Section~\ref{singlet_quant} gives examples of singlet quantities
both hadronic and gluonic (i.e.\ in this case $t_0$ and $w_0$)
and also discusses their $SU(3)$ flavour breaking expansions.
Section~\ref{latt_matt} first gives our lattice conventions,
ensembles used and numerical values of the singlet quantities.
This is followed by section~\ref{scale_set} in which the $\sqrt{t_0}$
and $w_0$ scales are determined for several lattice spacings.
In the next section, section~\ref{cont_res} we take
the continuum result to give the final result. Finally in
section~\ref{concl} we compare our result with other results for $n_f = 2+1$
flavours and give our conclusions.


\section{Extrapolating flavour singlet quantities}
\label{extrap_singlet}

\begin{figure}[!htbp]
\begin{center}

\begin{minipage}{0.45\textwidth}

   \begin{center}
      \includegraphics[width=6.50cm]{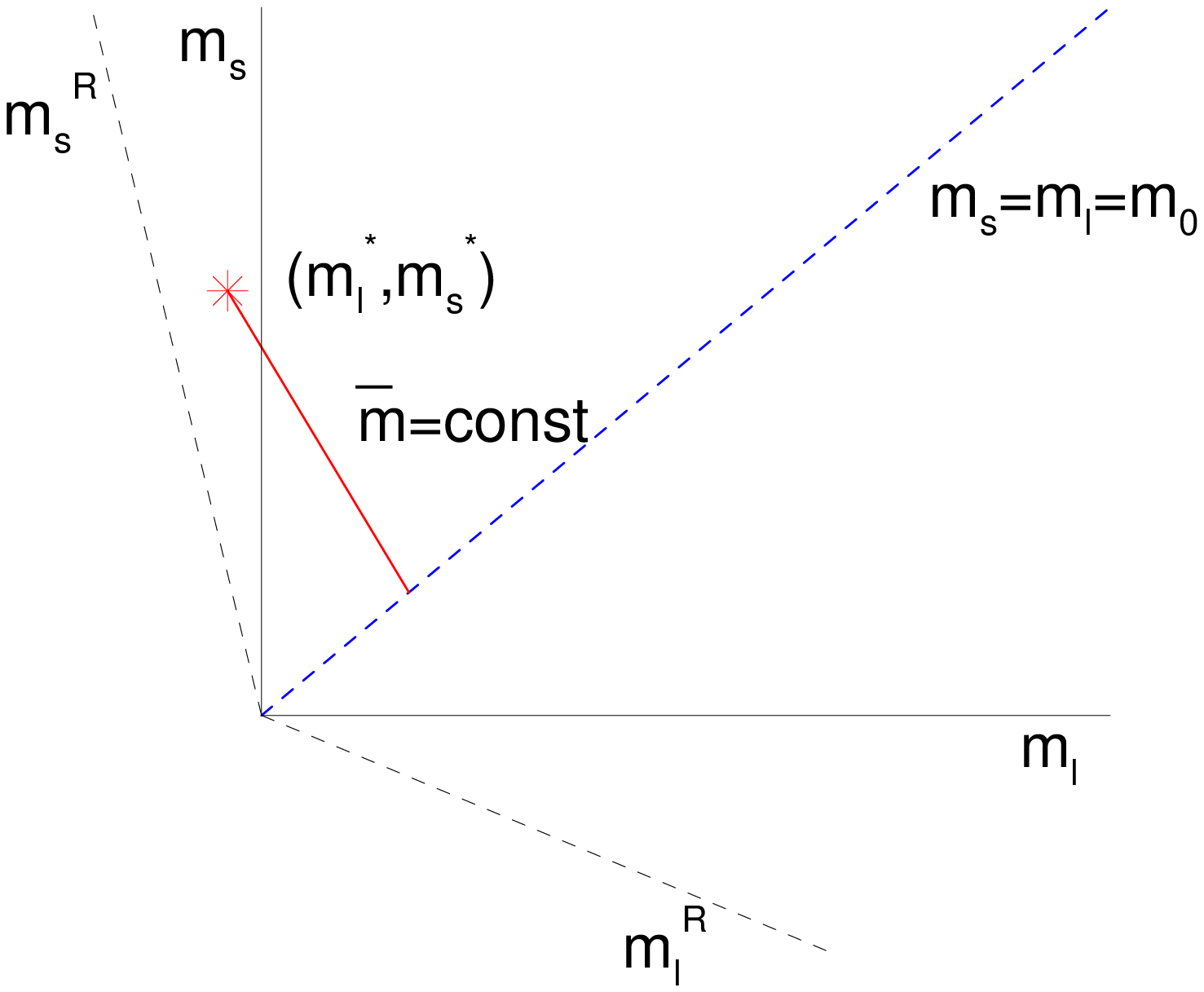}
   \end{center} 

\end{minipage}\hspace*{0.05\textwidth}
\begin{minipage}{0.45\textwidth}

   \begin{center}
      \includegraphics[width=6.50cm]{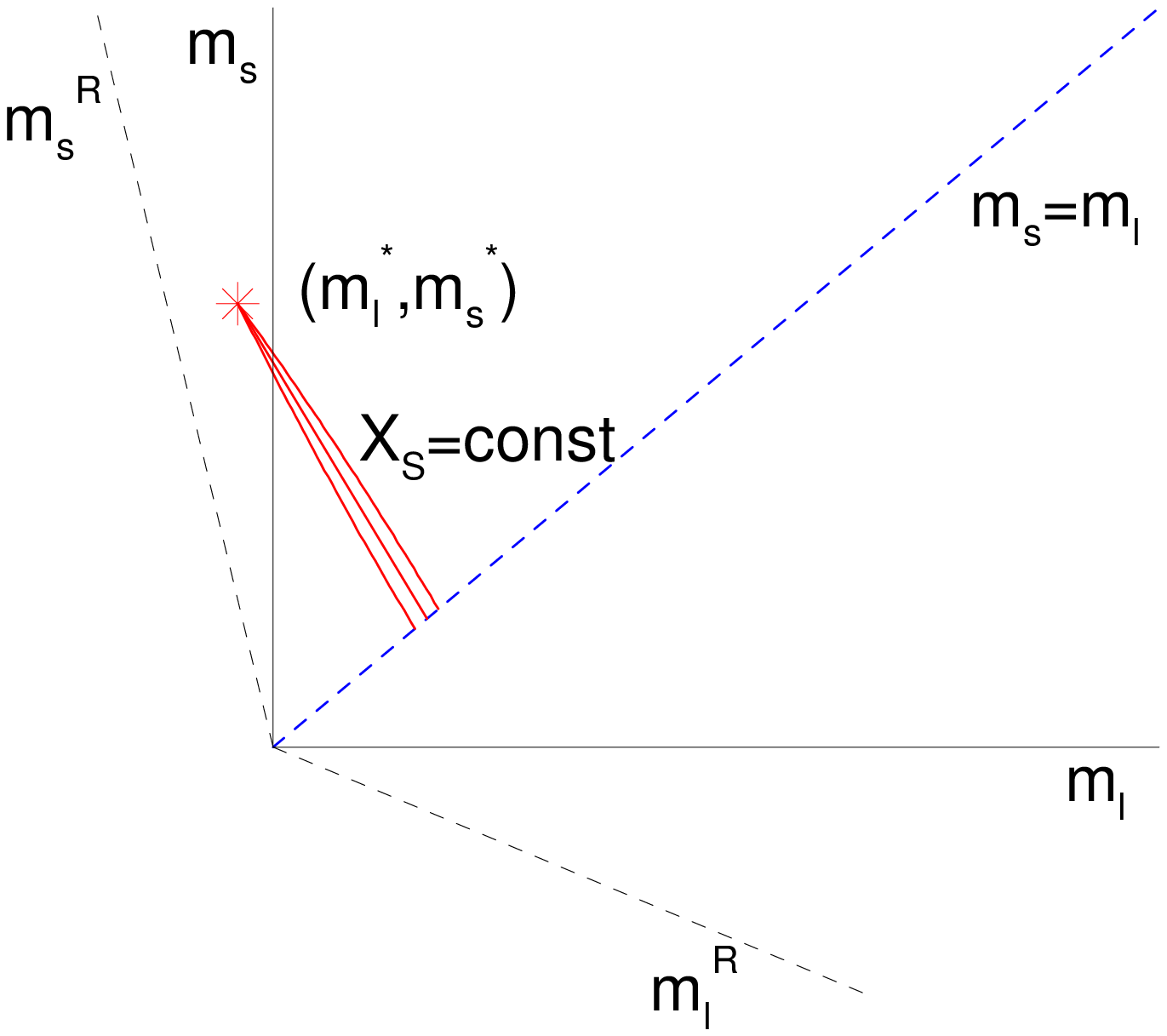}
   \end{center} 

\end{minipage}
\caption{Possible scenarios for the path in the $(m_u = m_d \equiv)\, m_l$
         -- $m_s$ plane. In the LH panel $\overline{m}=\mbox{const.}$,
         while in the RH panel we hold other singlet quantities constant.}
\label{sketch_paths}

\end{center}
\end{figure}

We consider extrapolations to the physical point from a point on the
$SU(3)$ flavour symmetric line keeping the average quark mass fixed,
\cite{bietenholz10a,bietenholz11a},
\begin{eqnarray}
   \overline{m} 
      = {\textstyle{1 \over 3}}(m_u + m_d + m_s) = \mbox{const.} \,.
\label{const_m}
\end{eqnarray}
This means that as the pion mass tends downwards to its physical value,
the kaon mass increases upwards to its physical value. (In particular
the kaon mass is never larger than its physical value.) 
Possible scenarios are sketched in Fig.~\ref{sketch_paths}
for a path from a point on the $SU(3)$ flavour symmetric line
to the physical point for the case of mass degenerate $u$ and $d$
quarks $m_l$, see eq.~(\ref{mass_degen}). (For non-mass degenerate $u$ and
$d$ quarks we would have instead a plane.) Shown in Fig.~\ref{sketch_paths}
are the bare and renormalised quark masses for the case discussed here 
of clover fermions. (Because the singlet and non-singlet pieces
renormalise differently the renormalised quark axes are not
orthogonal to each other, as further discussed in \cite{bietenholz11a}.
For chiral fermions this would not be the case.)
In the left hand panel one common trajectory is sufficient, while
in the right hand panel it depends on the singlet quantity
used, \cite{bietenholz11a}.

With the condition of eq.~(\ref{const_m}) flavour singlet quantities
turn out to have a stationary point in the quark mass starting from
a given point on the $SU(3)$ flavour symmetric line.
As we shall see this potentially allows simpler extrapolations
to the physical point. This property may be shown by considering
small changes about a given point on the $SU(3)$ flavour symmetric line.
Let $X_S(m_u, m_d, m_s)$ be a flavour singlet object i.e.\ $X_S$ is
invariant under the quark permutation symmetry between $u$, $d$ and $s$.
Then Taylor expanding $X_S$ about a point $m_0$ on the $SU(3)$ flavour
symmetric line $m_u = m_d = m_s = m_0 \equiv \overline{m}$,
\begin{eqnarray}
   m_q = \overline{m} + \delta m_q\,,
\label{dmq_def}
\end{eqnarray}
gives
\begin{eqnarray}
   \lefteqn{
      X_S( \overline{m} + \delta m_u,
           \overline{m} + \delta m_d,
           \overline{m} + \delta m_s )}
     & &                                                                \\
     &=& X_S(\overline{m}, \overline{m}, \overline{m})
      + \left. { \partial X_S \over \partial m_u } \right|_0 \delta m_u
      + \left. { \partial X_S \over \partial m_d } \right|_0 \delta m_d
      + \left. { \partial X_S \over \partial m_s } \right|_0 \delta m_s
      + O( (\delta m_q)^2 ) \,.
                                                             \nonumber
\end{eqnarray}
But on the symmetric line we have
\begin{equation}
   \left. { \partial X_S \over \partial m_u } \right|_0
      = \left. { \partial X_S \over \partial m_d } \right|_0
      = \left. { \partial X_S \over \partial m_s } \right|_0 \,,
\end{equation}
and on our chosen trajectory, eq.~(\ref{const_m})
\begin{eqnarray}
   \delta m_u + \delta m_d + \delta m_s = 0 \,,
\label{sum_dm}
\end{eqnarray}
which together imply that
\begin{eqnarray}
   X_S( \overline{m} + \delta m_u, \overline{m} + \delta m_d, 
        \overline{m} + \delta m_s )
     = X_S(\overline{m}, \overline{m}, \overline{m}) + O( (\delta m_q)^2 ) \,.
\label{Xs_expansion}
\end{eqnarray}
In other words, the effect at first order of changing the
strange quark mass is cancelled by the change in the light quark mass,
so we know that $X_S$ must have a stationary point on the $SU(3)$
flavour symmetric line.

Thus provided the quadratic terms in eq.~(\ref{Xs_expansion}) are
small it will not matter which singlet quantity we use, all are equivalent
to keeping $\overline{m} = \mbox{const.}$, the scenario of the left
hand panel of Fig.~\ref{sketch_paths}.


\section{Defining singlet quantities}
\label{singlet_quant}


We shall consider here hadronic singlet quantities, such as independent
averages with respect to quark permutations of the pseudoscalar,
vector and nucleon octets, \cite{bietenholz10a,bietenholz11a},
and also gluonic quantities derived from the Wilson flow,
\cite{luscher10a,borsanyi12a}. In particular we shall consider
\begin{eqnarray}
   X^2_\pi &=& \textstyle{1 \over 6}
                    (M_{K^+}^2 + M_{K^0}^2 + M_{\pi^+}^2 + M_{\pi^-}^2
                             + M_{K^-}^2 + M_{\bar{K}^0}^2)
             = \textstyle{1 \over 3}(2M_K^2 + M_\pi^2)
                                                             \nonumber \\
   X^2_\rho &=& \textstyle{1 \over 6}
                    (M_{K^{*+}}^2 + M_{K^{*0}}^2 + M_{\rho^+}^2 + M_{\rho^-}^2
                               + M_{\bar{K}^{*0}}^2 + M_{K^{*-}}^2)
              = \textstyle{1 \over 3}(2M_{K^*}^2 + M_\rho^2)
                                                             \nonumber \\
   X^2_N   &=& \textstyle{1 \over 6}
                    (M_p^2 + M_n^2 + M_{\Sigma^+}^2 + M_{\Sigma^-}^2
                           + M_{\Xi^0}^2 + M_{\Xi^-}^2)
              = \textstyle{1 \over 3}(M_N^2 + M_\Sigma^2 + M_\Xi^2)
                                                             \nonumber \\
   X^2_{t_0} &=& 1 / t_0
                                                             \nonumber \\
   X^2_{w_0} &=& 1 / w^2_0 \,.
\end{eqnarray}
(The charge conjugate mesons have the same masses, at least for pure QCD.
For example $M_{\pi^+} = M_{\pi^-}$, $M_{K^+} = M_{K^-}$ and
$M_{K^0} = M_{\overline{K}^0}$, obviously no such similar result holds for
the baryons.) The second expressions are the masses for mass degenerate
$u$ and $d$ quarks,
\begin{eqnarray}
   m_u = m_d \equiv m_l \,.
\label{mass_degen}
\end{eqnarray}
Except for $X_\pi^2$, which is naturally an average over quadratic masses,
it does not matter whether we consider linear or quadratic averages -- 
quadratic averages were found to give slightly better fits for heavy
partially quenched masses (up to the charm quark); in the small quark
mass range considered here this is less important. Note also that for the
Wilson flow singlets we consider the inverse of the flow variable
as then all singlet quantities have the same dimensions.

Other possibilities, as discussed in \cite{bietenholz11a} include
the further nucleon octet singlet
\begin{eqnarray}
   X^2_\Lambda
           &=& \textstyle{1 \over 2}
                    (M_{\Lambda^0}^2 + M_{\Sigma^0}^2)
              = \textstyle{1 \over 2}(M_\Lambda^2 + M_\Sigma^2) \,,
\end{eqnarray}
and baryon decuplet singlets
\begin{eqnarray}
   X^2_\Delta &=& \textstyle{1 \over 3}
                    (M_{\Delta^{++}}^2 + M_{\Delta^-}^2 + M_{\Omega^-}^2)
                 = \textstyle{1 \over 3}(2M_{\Delta}^2 + M_{\Omega}^2)
                                                             \nonumber \\
   X^2_{\Xi^*} &=& \textstyle{1 \over 6}
                    (M_{\Delta^+}^2 + M_{\Delta^0}^2
                                   + M_{\Sigma^{*+}}^2 + M_{\Sigma^{*-}}^2 
                                   + M_{\Xi^{*0}}^2 + M_{\Xi^{*-}}^2)
                 = \textstyle{1 \over 3}
                    (M_{\Delta}^2 + M_{\Sigma}^2 + M_{\Xi}^2)
                                                             \nonumber \\
   X_{\Sigma^*}^2 &=& M_{\Sigma^{*0}}^2 =  M_{\Sigma^*}^2 \,.
\end{eqnarray}
Further possibilities can be constructed from `fictitious' particles,
such as a `nucleon', $N_s$, with three mass degenerate quarks at the
strange quark mass. At the $SU(3)$ flavour symmetric points,
all the baryon octet hadrons are mass degenerate (at least from
a QCD perspective), so we expect that away from this point there
is no (or very little) difference between $X_N^2$ and $X_\Lambda^2$,
i.e.\ $X_N = X_\Lambda$. The same argument holds for the various
baryon decuplet possibilities, i.e.\ $X_\Delta = X_{\Xi^*} = X_{\Sigma^*}$.

How far does this extend? Let us consider the experimental (or
phenomenological) singlet hadron mass results%
\footnote{Only when necessary and for clarity  do we distinguish
between experimental and lattice masses -- $X_S^{\exp}$ and $X_S^{\rm lat}$
respectively.}
as given in Table~\ref{hadron_masses_singlet_avmass2}%
\footnote{As we are not considering mass differences, then the effect of 
electromagnetic effects is small, and so we can disregard them here.
For example for the lightest particles -- the pseudoscalar octet,
the value given in Table~\ref{hadron_masses_singlet_avmass2} for $X_\pi$
is to be compared with the value upon using Dashen's theorem which
gives $0.4116\,\mbox{GeV}$ (see e.g.\ \cite{horsley12a}). This
is a $\lsim 0.2\%$ difference. For the baryons for $X_N$ it is
$\lsim 0.1\%$ \label{emfoot}}.
\begin{table}[!t]
   \begin{center}
      \hspace*{-0.50in}
      \begin{tabular}{c|l}
         \hline
         \multicolumn{1}{c|}{Singlet quantity} & \multicolumn{1}{c}{GeV}\\
         \hline
         $X_\pi^{\exp}$     & 0.4126     \\  
         \hline
         $X_\rho^{\exp}$    & 0.8562     \\
         $X_N^{\exp}$       & 1.1610     \\
         $X_\Lambda^{\exp}$  & 1.1548     \\
         \hline
         $X_\Delta^{\exp}$   & 1.3944(12) \\
         $X_{\Xi^*}^{\exp}$  & 1.3837(1)  \\
         $X_{\Sigma^*}^{\exp}$& 1.3888(6)  \\
         \hline
     \end{tabular}
   \end{center}
\caption{Experimental values for various singlet quantities,
         averaging over $(\mbox{masses})^2$ (to $4$ decimal places).}
\label{hadron_masses_singlet_avmass2}
\end{table}
It is seen that even after the extrapolation from the $SU(3)$ flavour
line to the experimental point, then $X_N^{\exp} \approx X_\Lambda^{\exp}$
and $X_\Delta^{\exp} \approx X_{\Xi^*}^{\exp} \approx X_{\Sigma^*}^{\exp}$
the worst discrepancy (between  $X_{\Xi^*}^{\exp}$ and $X_{\Sigma^*}^{\exp}$)
is only a fraction of a percent. This indicates that quite likely
the $X_S$ are constant over a large interval.

However as the baryon decuplet possibilities are numerically substantially 
noisier, we shall only use the baryon octet hadrons here.

An equivalent statement (for singlet quantities built from hadron
masses) is found by considering the $SU(3)$ flavour breaking
expansion. As discussed in \cite{bietenholz10a,bietenholz11a} we have
for the octet mesons, $\pi^+(u\overline{d})$, $\pi^-(d\overline{u})$,
$K^+(u\overline{s})$, $K^-(s\overline{u})$, $K^0(d\overline{s})$
and $\overline{K}^0(s\overline{d})$ not lying at the centre of
the octet
\begin{eqnarray}
   M^2(a\overline{b}) = M_0^2 + \alpha_\pi(\delta m_a + \delta m_b) 
                             + O((\delta m_q)^2) \,,
\label{fit_mpsO}
\end{eqnarray}
where $a$ and $b$ are $u$, $d$ or $s$ quarks. Similar results
hold for the vector mesons.
For the $p(uud)$, $n(ddu)$, $\Sigma^+(uus)$, $\Sigma^-(dds)$, 
$\Xi^0(ssu)$, $\Xi^-(ssd)$ baryons on the
outer ring of the baryon octet%
\footnote{For non-degenerate $u$ and $d$ quark mass, the Lambda and
Sigma particles mix. This mixing is however very small. This 
was investigated in \cite{horsley14a}, where using the notation
there, it was shown that
$\textstyle{1 \over 2}(M_{\Lambda^0}^2 + M_{\Sigma^0}^2)
 = P_{A_1} = M_{0N}^2 + O((\delta m_q)^2)$, i.e.\ again we have
no linear term in the quark mass. Note also that when the $u$
and $d$ quark masses are degenerate, then no mixing occurs.}
we have
\begin{eqnarray}
   M^2(aab) = M_{0N}^2 + A_1(2\delta m_a + \delta m_b) 
                      + A_2(\delta m_b - \delta m_a) 
                      + O((\delta m_q)^2) \,.
\label{fit_mNO}
\end{eqnarray}
All the expansion coefficients are functions of $\overline{m}$.
It is easy to check that this means that
$X_S^2 = M_{0S}^2 + O((\delta m_l)^2)$, $S = \pi$, $\rho$ and $N$
in agreement with eq.~(\ref{Xs_expansion}).

`Fan' plots from the symmetric point down to the physical point
are well described by the linear behaviour of eqs.~(\ref{fit_mpsO})
and (\ref{fit_mNO}) as shown in 
\cite{bietenholz10a,bietenholz11a,horsley12a,horsley14a} which
further supports the earlier statement that $X_S$ is constant over
a large quark mass range.

Although in our extrapolations, we shall not be using chiral perturbation
theory, $\chi$PT, it is natural to ask about its relationship to the $SU(3)$
flavour breaking expansion. This also allows a check on 
eq.~(\ref{Xs_expansion}), assuming we are in a region where
$\chi$PT is valid. This was investigated in 
\cite{bietenholz10a,bietenholz11a} for hadron mass singlets 
and we now extend the argument to $t_0$ (and $w_0$),
using the result of \cite{bar13a}. Using the notation of this
paper and for mass degenerate $u$ and $d$ quark masses we find
\begin{eqnarray}
   t_0 = T(\overline{\chi}) \left[ 
          1 + {1 \over (4 \pi f_0)^4}
                ( \textstyle{5 \over 6}k_2 
                 + \textstyle{1 \over 4} k_5^{\prime\prime} )
                        (\chi_s - \chi_l)^2 + \cdots \right] \,,
\end{eqnarray}
where
\begin{eqnarray}
   T(\overline{\chi})
     = t_{\rm 0,ch} 
       \left[ 1 + {3 k_1 \over (4 \pi f_0)^2}\, \overline{\chi}
                + {8 k_2 \over (4 \pi f_0)^4}\,
                     \overline{\chi}^2 
                       \ln {\overline{\chi} \over \Lambda^2} 
                + {9 k_4^\prime \over (4 \pi f_0)^4}\, \overline{\chi}^2 \right] 
\end{eqnarray}
being the value of $t_0$ on the symmetric line ($t_{\rm 0,ch}$ is the
value in the chiral limit). $k_i$ are constants,
$f_0$ the pion decay constant again in the chiral limit and
$\chi_l = B_0m_l$, $\chi_s = B_0m_s$,
$\overline{\chi} = \textstyle{1 \over 3}(2\chi_l + \chi_s)$.

As expected, there is no linear term, and the first term we see 
is quadratic in the $SU(3)$ breaking. Further details are given 
in Appendix~\ref{chiPT}.


\section{Lattice matters}
\label{latt_matt}


\subsection{General}


We consider $2+1$ non-perturbatively $O(a)$ improved clover fermions,
as described in \cite{cundy09a}. The relation between the bare quark
masses $m_q$ in lattice units and the lattice mass parameters $\kappa_q$
is given by \cite{bietenholz10a,bietenholz11a}
\begin{eqnarray}
   m_q = {1 \over 2} 
            \left ({1\over \kappa_q} - {1\over \kappa_{0c}} \right)
            \qquad \mbox{with} \quad q \in \{l, s, 0\} \,,
\label{kappa_bare}
\end{eqnarray}
where in simulations we have mass degenerate $u$ and $d$ quarks,
i.e.\ $m_u = m_d \equiv m_l$ and the $s$ quark has mass $m_s$.
Along the $SU(3)$ flavour mass degenerate line, the common quark mass
is denoted by $m_0$ (or equivalently by $\kappa_0$) and where vanishing
of the quark mass along this line determines $\kappa_{0c}$. Along the
$\overline{m} = m_0 = \mbox{constant}$ line gives
from eqs.~(\ref{dmq_def}) and (\ref{kappa_bare}) the $SU(3)$
flavour breaking mass parameter as
\begin{equation}
   \delta m_q = {1 \over 2}
                 \left( {1 \over \kappa_q} - {1 \over \kappa_0}
                 \right) \,.
\label{delta_muq_kappa}
\end{equation}
We see that $\kappa_{0c}$ has dropped out of this equation,
so we do not need its explicit value here. Along this trajectory
the choice of quark masses is restricted and we have
\begin{eqnarray}
   \kappa_s = { 1 \over { {3 \over \kappa_0} - {2 \over \kappa_l} } } \,,
\label{kappas_mbar_const}
\end{eqnarray}
so once we have decided on a $\kappa_0$, then a given $\kappa_l$
determines $\kappa_s$. 

We consider four beta values $\beta = 5.8$, $5.65$, $5.50$, $5.40$
(where $\beta = 10/g_0^2$ with our conventions). In Tables~\ref{b5p80_params},
\ref{b5p65_params}, \ref{b5p50_params} and 
\ref{b5p40_params} we give parameters of the runs.
\begin{table}[!tb]
\begin{center}
   \begin{tabular}{ccc|cc|cc}
      $\beta$ &  $V$  & $\kappa_0$ & $\kappa_l$ & $\kappa_s$ & $M_\pi L$ & $L\,[\mbox{fm}]$ \\
      \hline

      5.80  &  $48^3\times 96$ &  0.122760  &  0.122760  &  0.122760  &  6.95 &  2.82 \\
      \hline
      5.80  &  $48^3\times 96$ &  0.122810  &  0.122810  &  0.122810  &  6.11 &  2.82 \\
      5.80  &  $48^3\times 96$ &  0.122810  &  0.122880  &  0.122670  &  5.11 &  2.82 \\
      5.80  &  $48^3\times 96$ &  0.122810  &  0.122940  &  0.122551  &  4.01 &  2.82 \\
      \hline
      5.80  &  $48^3\times 96$ &  0.122870  &  0.122870  &  0.122870  &  4.96 &  2.82 \\

   \end{tabular}
\end{center}
\caption{Parameters for $\beta = 5.80$. Each block has the same $\kappa_0$,
         i.e.\ constant $\overline{m}$.}
\label{b5p80_params}
\end{table}
\begin{table}[!tb]
\begin{center}
   \begin{tabular}{ccc|cc|cc}
      $\beta$ &  $V$  & $\kappa_0$ & $\kappa_l$ & $\kappa_s$ & $M_\pi L$ & $L\,[\mbox{fm}]$ \\
      \hline

      5.65  &  $32^3\times 64$ &  0.121975  &  0.121975  &  0.121975  &  4.99 &  2.19 \\
      \hline
      5.65  &  $32^3\times 64$ &  0.122005  &  0.122005  &  0.122005  &  4.67 &  2.19 \\
      5.65  &  $32^3\times 64$ &  0.122005  &  0.122078  &  0.121859  &  4.00 &  2.19 \\
      {\it 5.65}  &  {\it $32^3\times 64$} &  {\it 0.122005}  &  {\it 0.122130}  &  {\it 0.121756}  &  {\it 3.44} &  {\it 2.19} \\
      \hline
      5.65  &  $32^3\times 64$ &  0.122030  &  0.122030  &  0.122030  &  4.32 &  2.19 \\
      \hline
      5.65  &  $32^3\times 64$ &  0.122050  &  0.122050  &  0.122050  &  4.03 &  2.19 \\

   \end{tabular}
\end{center}
\caption{Parameters for $\beta = 5.65$.
         The entries in italics have $M_\pi L < 4$.}
\label{b5p65_params}
\end{table}
\begin{table}[!tb]
\begin{center}
   \begin{tabular}{ccc|cc|cc}
      $\beta$ &  $V$  & $\kappa_0$ & $\kappa_l$ & $\kappa_s$ & $M_\pi L$ & $L\,[\mbox{fm}]$ \\
      \hline

      5.50  &  $32^3\times 64$ &  0.120900  &  0.120900  &  0.120900  &  5.59 &  2.37 \\
      5.50  &  $32^3\times 64$ &  0.120900  &  0.121040  &  0.120620  &  4.32 &  2.37 \\
      {\it 5.50}  &  {\it $32^3\times 64$} &  {\it 0.120900}  &  {\it 0.121095}  &  {\it 0.120512}  &  {\it 3.72} &  {\it 2.37} \\
      {\it 5.50}  &  {\it $32^3\times 64$} &  {\it 0.120900}  &  {\it 0.121145}  &  {\it 0.120413}  &  {\it 3.10} &  {\it 2.37} \\
      5.50  &  $48^3\times 96$ &  0.120900  &  0.121166  &  0.120371  &  4.10 &  3.55 \\
      \hline
      5.50  &  $32^3\times 64$ &  0.120920  &  0.120920  &  0.120920  &  5.27 &  2.37 \\
      5.50  &  $32^3\times 64$ &  0.120920  &  0.121050  &  0.120661  &  4.10 &  2.37 \\
      \hline
      5.50  &  $32^3\times 64$ &  0.120950  &  0.120950  &  0.120950  &  4.83 &  2.37 \\
      5.50  &  $32^3\times 64$ &  0.120950  &  0.121040  &  0.120770  &  3.97 &  2.37 \\
      {\it 5.50}  &  {\it $32^3\times 64$} &  {\it 0.120950}  &  {\it 0.121099}  &  {\it 0.120653}  &  {\it 3.24} &  {\it 2.37} \\
      \hline
      5.50  &  $32^3\times 64$ &  0.120990  &  0.120990  &  0.120990  &  4.11 &  2.37 \\

   \end{tabular}
\end{center}
\caption{Parameters for $\beta = 5.50$.}
\label{b5p50_params}
\end{table}
\begin{table}[!tb]
\begin{center}
   \begin{tabular}{ccc|cc|cc}
      $\beta$ &  $V$  & $\kappa_0$ & $\kappa_l$ & $\kappa_s$ & $M_\pi L$ & $L\,[\mbox{fm}]$ \\
      \hline

      5.40  &  $24^3\times 48$ &  0.119860  &  0.119860  &  0.119860  & 4.98 & 1.96 \\
      \hline
      5.40  &  $24^3\times 48$ &  0.119895  &  0.119895  &  0.119895  & 4.54 & 1.96 \\
      \hline
      5.40  &  $24^3\times 48$ &  0.119930  &  0.119930  &  0.119930  & 4.11 & 1.96 \\
      {\it 5.40}  &  {\it $24^3\times 48$} &  {\it 0.119930}  &  {\it 0.120048}  &  {\it 0.119695}  & {\it 3.35} & {\it 1.96} \\
      \hline
      {\it 5.40}  &  {\it $24^3\times 48$} &  {\it 0.120000}  &  {\it 0.120000}  &  {\it 0.120000}  & {\it 3.25} & {\it 1.96} \\

   \end{tabular}
\end{center}
\caption{Parameters for $\beta = 5.40$.}
\label{b5p40_params}
\end{table}
The entries in italics have $M_\pi L < 4$.

For example we consider $\beta = 5.50$ on $32^3\times 64$ lattices
with degenerate quark masses of $\kappa_0 = 0.120900$, $0.120920$,
$0.120950$ and $0.120990$ which, as we will see, encompasses
the initial $SU(3)$ flavour symmetric point on the constant
$\overline{m}$ trajectory to the physical point. As can be seen for
some of the $\kappa_0$ values we have extended the constant
$\overline{m} = m_0$ trajectories down in the direction of the physical point.

While changing the $\beta$ value gives the greatest change to the
singlet terms, smaller effects occur on the $SU(3)$ flavour symmetric line
as we change $m_0$. This will help us to locate the initial
$\kappa_0$, the starting point for the trajectory in the $m_s$--$m_l$
plane leading to the physical point. Note that if $\overline{m}$ is held
constant, then a further advantage of this condition is that for
clover fermions the $O(a)$ improved coupling constant remains unchanged as
\begin{eqnarray}
   \tilde{g}_0^2 = g_0^2( 1 + b_g(g_0)a\overline{m}) \,,
\end{eqnarray}
although this is unlikely to lead to any large effect.

For orientation the $SU(3)$ symmetric point has a pion mass of
about $\sim 450\,\mbox{MeV}$ and we reach down to about $\sim 260\,\mbox{MeV}$.

The specific components used in the flow discretisation here are%
\footnote{The flow Wilson action here means
$S_{\rm flow}[U] = \sum {\rm Re\,Tr} [ 1 - U^{\rm plaq} ]$.}
\begin{eqnarray}
   \lefteqn{\left( \rm f[low], g[auge \, action], o[bservable] \right)}
                  \hspace*{1.5in}
     & &                                                 \nonumber  \\
     &=& \left( \rm W[ilson], S[ymanzik\,[tree\, level]], C[lover] \right) \,,
\end{eqnarray}
and the Runge-Kutta discretisation is used for the
flow equation, \cite{luscher10a}.

One can improve the scaling behaviour, which is expected to
have $O(a^2)$ corrections. For example for $\sqrt{t_0}$
following \cite{bazavov15a} we can write
\begin{eqnarray}
   \left. {F(t) \over 1 + C_2{a^2\over t} + \ldots}
                                \right|_{t=t_{0\,\rm imp}(c)} = c
   \quad \Rightarrow \quad
   t_{0\,\rm imp} = t_0\,\left( 1 + C_2{F_0 \over F_0^\prime}\,
                                {a^2 \over t_0} + \ldots \right) \,.
\end{eqnarray}
Inverting this and relabelling $t_{0\,\rm imp} \to t_{0\,\rm cont}$ gives
\begin{eqnarray}
   t_0 = t_{0\,\rm cont} 
           \left( 1 - C_2{F_{0\,\rm cont} \over  F_{0\,\rm cont}^\prime} \, 
                     {a^2 \over t_{0\,\rm cont}} + \ldots 
            \right) \,,
\label{a2_expan}
\end{eqnarray}
where $F_x = F(t_x)$, $F^\prime_x = t d F(t)/dt |_{t_x}$, 
with $x \equiv 0$ or $0\,\rm cont$. At tree level for the case here
$(fgo) = (WSC)$, $C_2 = - 7/72$ \cite{fodor14a} so we expect the gradient
to be $+$ve.


\subsection{Singlet quantities}

\begin{figure}[!htb]
   \begin{center}
      \includegraphics[width=9.50cm]
         {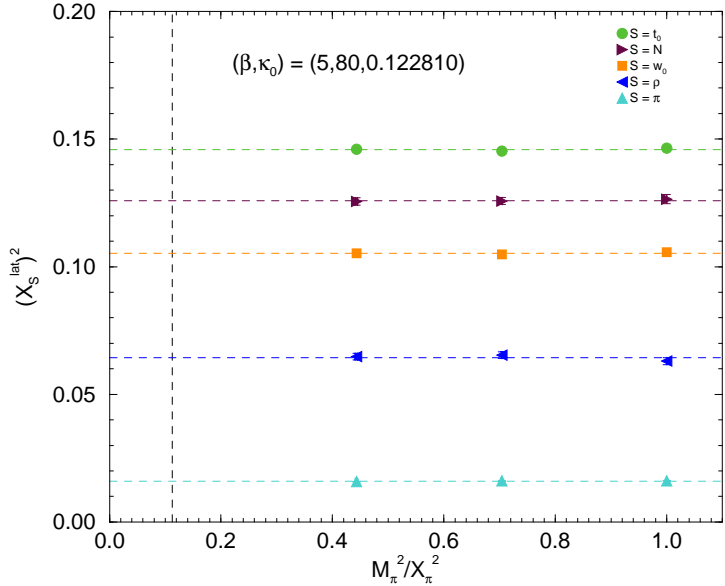}
   \end{center}
\caption{Top to bottom $(X_S^{\rm lat})^2$ for $S = t_0$ (circles),
         $N$ (right triangles), $w_0$ (squares), $\rho$ (left triangles)
         and $\pi$ (up triangles) for $(\beta, \kappa_0) = (5.80, 0.122810)$
         with constant fits.}
\label{b5p80_mbar_k0p122810_mpsO2o2mpsK2+mps2o3_aX2_48x96}
\end{figure}
\begin{figure}[!htb]
   \begin{center}
      \includegraphics[width=9.50cm]
         {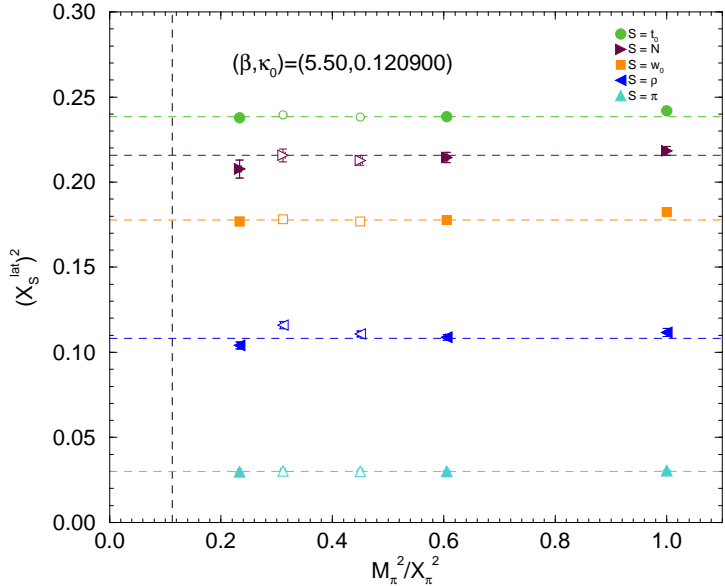}
   \end{center}
\caption{$(X_S^{\rm lat})^2$ for $S = t_0$, $N$, $w_0$, $\rho$ and $\pi$,
         (top to bottom) for $(\beta, \kappa_0) = (5.50, 0.120900)$
         together with constant fits. Same notation as for
         Fig.~\protect\ref{b5p80_mbar_k0p122810_mpsO2o2mpsK2+mps2o3_aX2_48x96}.
         Opaque points are not included in the fits, as they have $M_\pi L < 4$.}
\label{b5p50_mbar_k0p120900_mpsO2o2mpsK2+mps2o3_aX2_32x64+48x96}
\end{figure}

We first investigate the constancy of the singlet quantities $X_S^2$,
as discussed in section~\ref{extrap_singlet}.
In Fig.~\ref{b5p80_mbar_k0p122810_mpsO2o2mpsK2+mps2o3_aX2_48x96}
and \ref{b5p50_mbar_k0p120900_mpsO2o2mpsK2+mps2o3_aX2_32x64+48x96} we
plot $X_S^2$ for $S = t_0$, $N$, $w_0$ $\rho$ and $\pi$
against $M_\pi^2/X_\pi^2$ (which is equivalent to $1/\kappa_l$).
The value at $M_\pi^2/X_\pi^2 = 1$ corresponds
to the $SU(3)$ symmetric point and the vertical dashed lines
correspond to the physical point. 

No structure or trend is seen in the results and they are
compatible with $(X_S^{\rm lat})^2$ (and hence $X_S^{\rm lat}$) a constant
down to the vicinity of the physical point. The constant fits exclude
points with $M_\pi L < 4$. However these additional points all
have $M_\pi L > 3$ and are completely consistent with the fitted
points. Thus the constancy of $X_S$ as discussed in
sections~\ref{extrap_singlet} and \ref{singlet_quant} is
supported by the numerical results.

Furthermore any (significant) quadratic term would mean that
each quantity $S$ starts at a slightly different point on the
$SU(3)$ flavour symmetric line (but all with the same gradient)
and the trajectories would then all focus at the experimental point.
We have considered the partially quenched expansion (up to 
cubic terms in the quark mass) and have determined that along
the unitary line considered here these higher order terms
are negligible -- only when the quark mass is in the vicinity
of the charm quark mass do these non-linear terms become appreciable.
Thus practically we have a unique starting point on the
$SU(3)$ flavour symmetric line for the trajectory to the
physical point, i.e.\ we have the situation for at least $S = t_0$, $N$,
$w_0$, $\rho$ and $\pi$ of the left panel of Fig.~\ref{sketch_paths}.
To check this an alternative description is provided by a re-arrangement
of $X_\pi^2 / X_S^2 = (2M_K^2 + M_\pi^2) /  X_S^2$ to give 
\begin{eqnarray}
   {2M_K^2 - M_\pi^2 \over X^2_S} 
       = {X_\pi^2 \over X_S^2} - 2{M_\pi^2 \over X_S^2}\,,
\label{qm_plane}
\end{eqnarray}
for $S = N, \rho, t_0, w_0$. So plotting $(2M_K^2 - M_\pi^2) / X^2_S$
against $M_\pi^2 / X_S^2$ with constant gradient $-2$ should describe
the data for all $S$. Hence in Fig.~\ref{sketch_paths} left panel the
gradient is $-2$ while for the right panel, we would have an initial
gradient of $-2$ and then some curvature, but all meeting at
the physical point.

\begin{figure}[!htbp]
   \begin{center}
      \includegraphics[width=9.50cm]
         {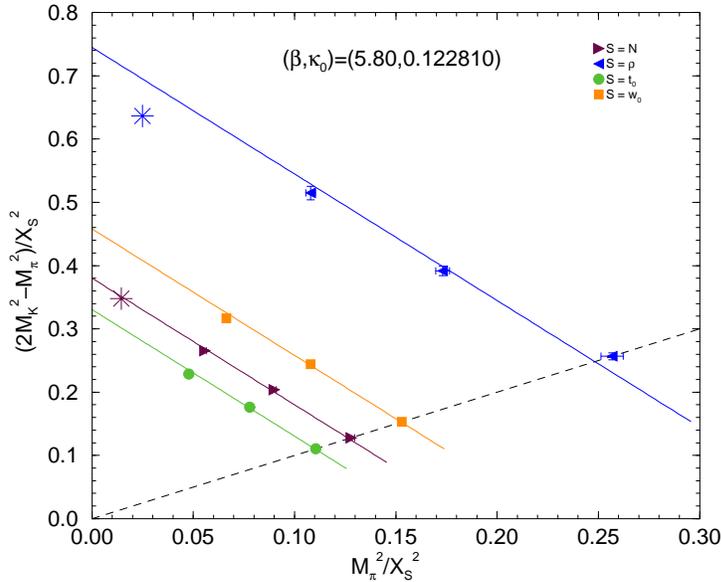}
   \end{center}
\caption{$(2M_K^2 - M_\pi^2) / X^2_S$ against $M_\pi^2 / X_S^2$,
         together with the fit from eq.~(\protect\ref{qm_plane})
         for $(\beta, \kappa_0) = (5.80, 0.122810)$ and
         $S = N$ (right triangles), $\rho$ (left triangles),
         $t_0$ (circles) and $w_0$ (squares). The stars correspond
         to the experimental values for $S = \rho$, $N$ upper and
         lower respectively.}
\label{b5p80_mps2oXS2_2mpsK2-mps2oXS2}
\end{figure}

\begin{figure}[!htbp]
   \begin{center}
      \includegraphics[width=9.50cm]
         {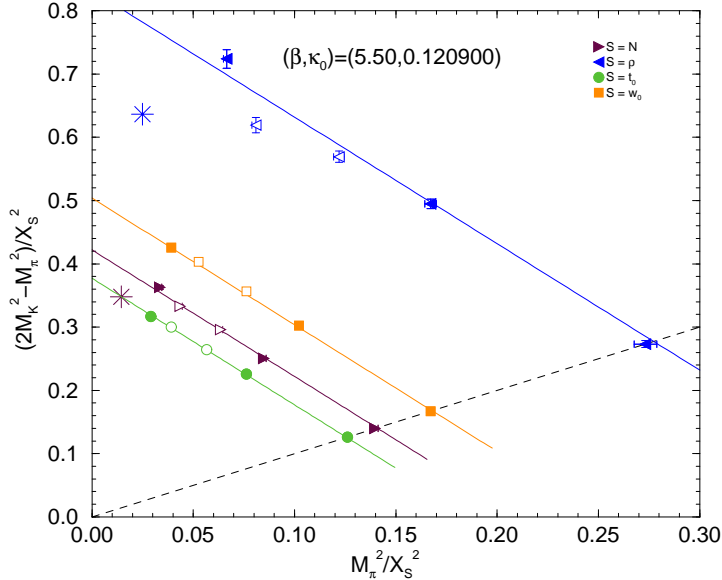}
   \end{center}
\caption{$(2M_K^2 - M_\pi^2) / X^2_S$ against $M_\pi^2 / X_S^2$,
         together with the fit from eq.~(\protect\ref{qm_plane})
         for $(\beta, \kappa_0) = (5.50, 0.120900)$ and
         $S = N$, $\rho$, $t_0$, $w_0$. The notation is as for
         Fig.~\protect\ref{b5p80_mps2oXS2_2mpsK2-mps2oXS2}.
         Opaque points are not included in the fit as they
         have $M_\pi L < 4$.}
\label{b5p50_mps2oXS2_2mpsK2-mps2oXS2}
\end{figure}

In Figs.~\ref{b5p80_mps2oXS2_2mpsK2-mps2oXS2} and 
\ref{b5p50_mps2oXS2_2mpsK2-mps2oXS2} we present fits of
eq.~(\ref{qm_plane}) to numerical results of $(2M_K^2 - M_\pi^2)/X_S^2$
versus $M_\pi^2/X_S^2$ with $S = N$, $\rho$, $t_0$, $w_0$ for the same data
sets as in Fig.~\ref{b5p80_mbar_k0p122810_mpsO2o2mpsK2+mps2o3_aX2_48x96}
and \ref{b5p50_mbar_k0p120900_mpsO2o2mpsK2+mps2o3_aX2_32x64+48x96}.
Again straight lines (from the fit function) describe the data
very well. However as can be seen the lines for the $S = N$ and $\rho$
cases do not quite go through their physical points (denoted by 
stars). This is because the $\kappa_0$ used while close is
not quite the value required for the correct path.
We shall in future denote this point by $\kappa_0^*$.
For example, it can be seen that the $\beta = 5.80$, $\kappa_0 = 0.12281$
lines are closer to the physical point, i.e.\ $\kappa_0 = 0.122810$
is closer to $\kappa_0^*$ than for the $\beta = 5.50$,
$\kappa_0 = 0.120900$ data. Again the picture is best described
by the left panel of Fig.~\ref{sketch_paths}. So at least for
all these quantities we only have to slightly tune to find the
appropriate $\kappa_0^*$ giving the beginning of the path from the
$SU(3)$ flavour symmetric line to the physical point.


\section{Scale setting}
\label{scale_set}


Using these results we now take $X_S = \mbox{const.}$ to
define the scale, i.e.\ we set
\begin{eqnarray}
   X_S^{\rm lat} = \mbox{const.} = a_SX_S^{\exp} \,.
\end{eqnarray}
We take the experimental hadron mass results as given in
Table~\ref{hadron_masses_singlet_avmass2}.
As $X_{t_0}$, $X_{w_0}$ are secondary quantities, i.e.\
$X_{t_0}^{\exp}$, $X_{w_0}^{\exp}$ are not experimentally known,
they have to be determined. 

If we now normalise
\begin{eqnarray}
   a^2_S(\kappa_0) 
      = { (X_S^{\rm lat}(\kappa_0))^2 \over (X_S^{\exp})^2 } \,,
\label{a2kapp0S}
\end{eqnarray}
this provides an estimate for the lattice spacing using the
singlet quantity $S$, which is also a function of the point
on the $SU(3)$ flavour symmetric line, i.e.\ $\kappa_0$
(we have indicated this by writing $a_S(\kappa_0)$ for the
lattice spacing).

We now vary $\kappa_0$, searching for the location where
the various $a_S(\kappa_0)$ cross, providing a value for
the common lattice spacing $a$ (and $\kappa_0^*$).
While ideally we would wish the crossing of all the lines
to occur at a single point leading to a common lattice spacing,
this, of course, does not quite happen. So we consider
pairs of singlet quantities and determine the crossing points,
together with the associated (bootstrap) error.
In particular we apply this to the pairs
\begin{eqnarray}
   (\pi,N), \, (\pi,\rho) \,.
\end{eqnarray}

We now use these crossings to adjust $X_{t_0}$ and $X_{w_0}$
so they also go through these points. This determines
$\sqrt{t_0^{\exp}}$, $w_0^{\exp}$. For example we have
\begin{eqnarray}
   (w_0^{\exp})^2 \equiv {1 \over (X_{w_0}^{\exp})^2}
          = { a^2 \over (X_{w_0}^{\rm lat})^2} \,.
\label{w0_determin}
\end{eqnarray}
 
\begin{figure}[!htbp]
   \begin{center}
      \includegraphics[width=9.50cm]
         {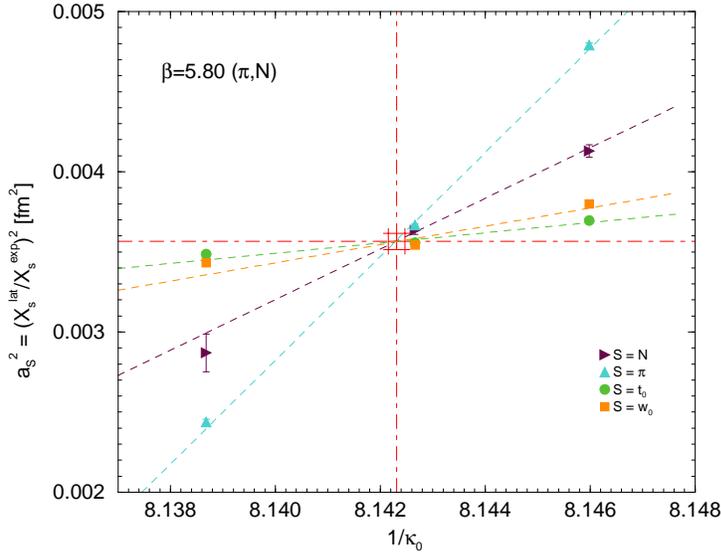}
   \end{center}
\caption{$a_S^2$ against $1/\kappa_0$ for $S = \pi$ (up triangles),
         $N$ (right triangles) and $t_0$ (circles), $w_0$ (squares)
         together with linear fits for $\beta = 5.80$.}
\label{b5p80_a2X2oX2expt_piN_sqrtt0+w0}
\end{figure}

\begin{figure}[!htbp]
   \begin{center}
      \includegraphics[width=9.50cm]
         {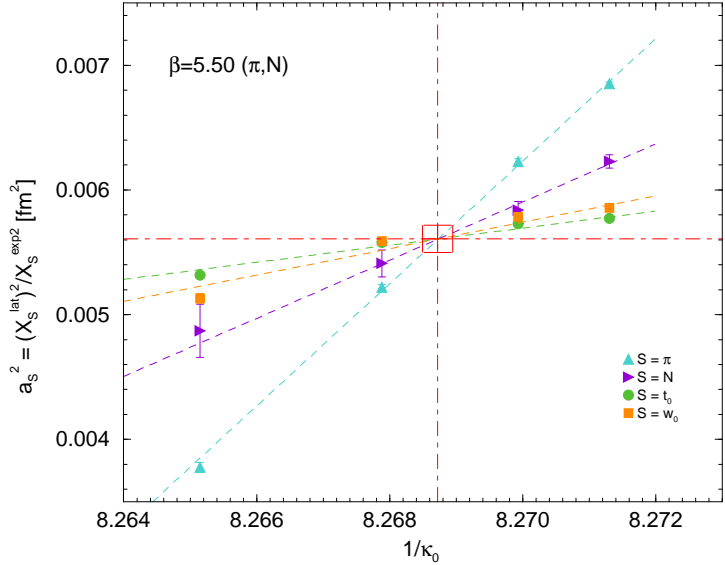}
   \end{center}
\caption{$a_S^2$ against $1/\kappa_0$ for $S = \pi$, $N$ and
         $t_0$, $w_0$ together with quadratic fits
         for $\beta = 5.50$. Notation as for
         Fig.~\protect\ref{b5p80_a2X2oX2expt_piN_sqrtt0+w0}.}
\label{b5p50_a2X2oX2expt_piN_sqrtt0+w0}
\end{figure}

For example in Figs.~\ref{b5p80_a2X2oX2expt_piN_sqrtt0+w0} and
\ref{b5p50_a2X2oX2expt_piN_sqrtt0+w0}
we plot $a_S^2(\kappa_0)$ from eq.~(\ref{a2kapp0S})
(in $\mbox{fm}^2$) against $\/\kappa_0$ for
$\beta = 5.80$ and $5.50$ for the singlet quantities
$S = \pi$ and $N$ together with $S = t_0$ and $w_0$.
Where we have three $\kappa_0$ values a linear fit in
$1 / \kappa_0$ is made, while if there are four $\kappa_0$
values available then a quadratic fit is made.
(However it made very little difference to the later results
whether the results from the linear or quadratic fit is used,
as mainly interpolations between the $X_S^{\rm lat}$ data is sufficient.)
Also plotted is $S = t_0$ and $w_0$, again together with
appropriate fits. The lattice values have been adjusted 
with a common factor so these singlet quantities also
cross at the same value as $(\pi, N)$, which is equivalent
to a determination of $\sqrt{t_0}^{\exp}$ and $w_0^{\exp}$
as indicated in eq.~(\ref{w0_determin}).
This procedure is then repeated for the pair $(\pi, \rho)$.

For completeness we also take a weighted average of both
the $(\pi,N)$ and $(\pi,\rho)$ crossings
to determine the best $(1/\kappa_0^*, a^2)$.
These values are given in Table~\ref{kapp0+a2_values}.
\begin{table}[!h]
   \begin{center}
      \begin{tabular}{c|rr|rr}
         \hline
         \multicolumn{1}{c|}{$\beta$} & 
         \multicolumn{1}{c}{$1/\kappa_0^*$} & 
         \multicolumn{1}{c|}{$a^2\,[\mbox{fm}^2$]} & 
         \multicolumn{1}{c}{$\kappa_0^*$} & 
         \multicolumn{1}{c}{$a\,[\mbox{fm}]$}   \\
         \hline

         $5.80$  & 8.14197(12) & 0.00346(04) & 0.122820(2) & 0.0588(03)  \\
         $5.65$  & 8.19602(15) & 0.00468(06) & 0.122010(2) & 0.0684(04)  \\
         $5.50$  & 8.26844(13) & 0.00547(06) & 0.120942(2) & 0.0740(04)  \\ 
         $5.40$  & 8.33823(25) & 0.00669(16) & 0.119930(4) & 0.0818(09)  \\

         \hline
     \end{tabular}
   \end{center}
\caption{Determined values of $1/\kappa_0^*$ and $a^2\,[\mbox{fm}^2]$.
         For completeness in the third and fourth columns we also give
         $\kappa_0^*$ and $a\,[\mbox{fm}]$ directly.}
\label{kapp0+a2_values}
\end{table}


\section{Continuum results}
\label{cont_res}

\begin{figure}[!htbp]
   \begin{center}
      \includegraphics[width=9.50cm]
         {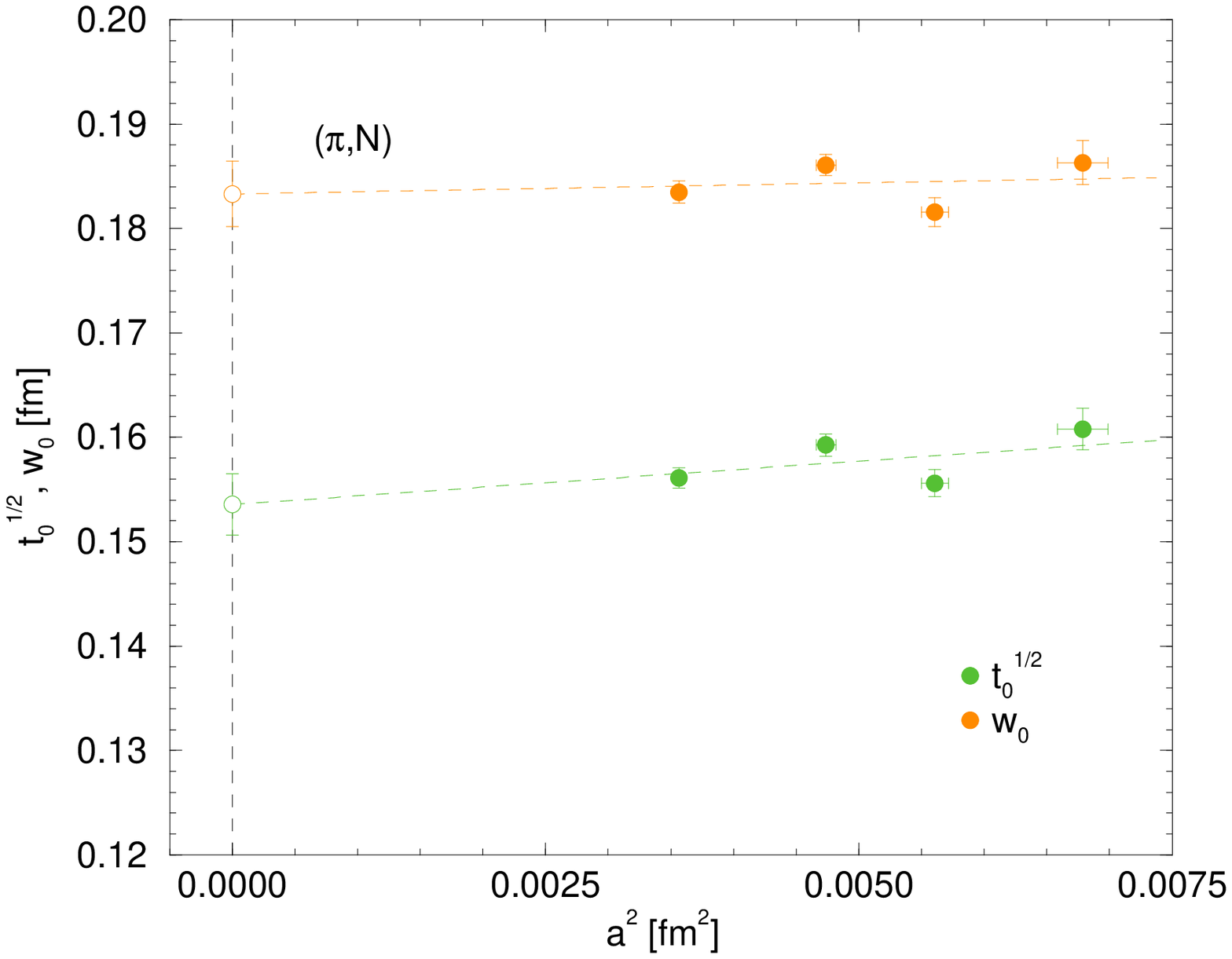}
   \end{center}
\caption{$\sqrt{t_0}$ and $w_0$ (in $\mbox{fm}$) against $a^2$
         (in $\mbox{fm}^2$) from the $(\pi, N)$ crossing together
         with a linear fit.}
\label{a2_sqrtt0+w0_N+pi}
\end{figure}
\begin{figure}[!htbp]
   \begin{center}
      \includegraphics[width=9.50cm]
         {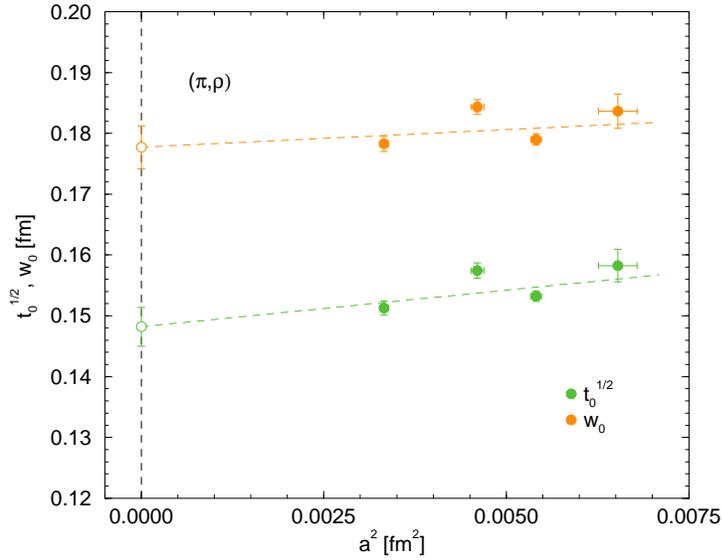}
   \end{center}
\caption{$\sqrt{t_0}$ and $w_0$ (in $\mbox{fm}$) against $a^2$
         (in $\mbox{fm}^2$) from the $(\pi, \rho)$ crossing together
         with a linear fit.}
\label{a2_sqrtt0+w0_rho+pi}
\end{figure}

We are now in a position to perform the last, continuum, extrapolation.
In Figs.~\ref{a2_sqrtt0+w0_N+pi}, \ref{a2_sqrtt0+w0_rho+pi}
we show these extrapolations from the pairs $(\pi, N)$ and $(\pi, \rho)$.
As anticipated the gradients in $a^2$, while small, are positive
(c.f.\ eq.~(\ref{a2_expan})) with the $(\pi,\rho)$ results being slightly
larger than the $(\pi,N)$ results.

Finally a weighted average from these continuum results
(i.e.\ for $a^2 = 0$) gives our final results
\begin{eqnarray}
   \sqrt{t_0^{\exp}} = 0.1511(22)(06)(05)(03)\,\mbox{fm}\,, \qquad
          w_0^{\exp} = 0.1808(23)(05)(06)(04)\,\mbox{fm} \,.
\label{phys_vals}
\end{eqnarray}
The first error is statistical, while the second (finite volume),
the third ($SU(3)$ flavour breaking expansion) and
fourth (scale) are systematic errors as discussed in
Appendix~\ref{systematic}.


\section{Conclusions}
\label{concl}


In this article we have described a method for determining the trajectory
to approach the physical point, and demonstrated (theoretically and
numerically) that singlet quantities remain constant as we approach this point.
This enables us by considering pairs of singlet quantities to
determine `best' lattice spacings and starting values for the path.
By matching these results to the flow variables $t_0$ and $w_0$
this enables a determination of their physical values,
see eq.~(\ref{phys_vals}).

\begin{figure}[!htbp]
   \begin{center}
      \includegraphics[width=9.50cm]
         {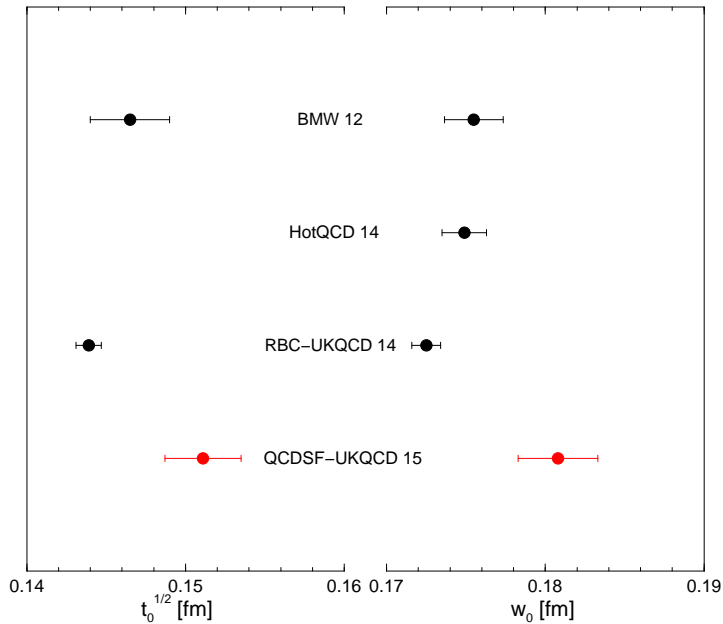}
   \end{center}
\caption{$\sqrt{t_0^{\exp}}$, left panel and $w_0^{\exp}$, right panel
         in $\mbox{fm}$ for BMW 12 \protect\cite{borsanyi12a},
         HotQCD 14 \protect\cite{bazavov14a}, 
         RBC-UKQCD 14 \protect\cite{blum14a}, together with the
         present results.}
\label{comparison_nf2p1}
\end{figure}

In Fig.~\ref{comparison_nf2p1} we compare these results
with other determinations for $n_f = 2+1$ flavours, namely
BMW 12 \cite{borsanyi12a}, HotQCD 14 \cite{bazavov14a}
and RBC-UKQCD 14 \protect\cite{blum14a}.
(The given errors are taken in quadrature.) Reasonable consistency
is found between the different determinations.

In conclusion we have determined in this article the flow scales for $t_0$
and $w_0$. These are `secondary' scales and while having the advantage
of being cheap and accurate to determine from lattice simulations
are not directly experimentally accessible and thus have to be matched
to physical quantities.


\section*{Acknowledgements}


The numerical configuration generation (using the BQCD lattice
QCD program \cite{nakamura10a}) and data analysis 
(using the Chroma software library \cite{edwards04a}) was carried out
on the IBM BlueGene/Q using DIRAC 2 resources (EPCC, Edinburgh, UK),
the BlueGene/P and Q at NIC (J\"ulich, Germany),
the Lomonosov at MSU (Moscow, Russia) and the
SGI ICE 8200 and Cray XC30 at HLRN (The North-German Supercomputer
Alliance) and on the NCI National Facility in Canberra, Australia
(supported by the Australian Commonwealth Government).
HP was supported by DFG Grant No. SCHI 422/10-1.
PELR was supported in part by the STFC under contract ST/G00062X/1
and JMZ was supported by the Australian Research Council Grant
No. FT100100005 and DP140103067. We thank all funding agencies.



\appendix

\section*{Appendix}


\section{Singlet chiral perturbation theory: Wilson flow}
\label{chiPT}


We want to check that the chiral perturbation theory 
results for the Wilson flow, as given in \cite{bar13a}
are consistent with the $SU(3)$ flavour symmetry expansion
\cite{bietenholz11a}.


\subsection{Pseudoscalar Meson masses}


First we need to set out some notation. The quark masses for the $2+1$
case are best denoted by $\chi_q$, defined through
\begin{eqnarray} 
    \chi_l &\equiv & B_0 m_l
                                                   \nonumber           \\ 
    \chi_s &\equiv & B_0 m_s \,.
\end{eqnarray}
To simplify expressions, it is useful to define some additional 
$\chi$ variables: 
\begin{eqnarray} 
   \overline{\chi} &\equiv & {\textstyle {1 \over 3}} (2 \chi_l + \chi_s)
                                                   \nonumber           \\ 
   \chi_\pi         &\equiv& \chi_l 
                                                   \nonumber           \\
   \chi_K          &\equiv& {\textstyle {1 \over 2}} (\chi_s + \chi_l) 
                                                   \nonumber           \\
   \chi_\eta        &\equiv& {\textstyle {1 \over 3}} (2 \chi_s + \chi_l) \,,
\end{eqnarray}
and a logarithmic function
\begin{eqnarray}
   \mu_P \equiv {\chi_P \over (4 \pi f_0)^2} \ln {\chi_P \over \Lambda^2}
         \approx {M^2_P \over (4 \pi f_0)^2} \ln {M^2_P \over \Lambda^2}\,, 
                  \qquad P \in \pi, K, \eta \,.
\end{eqnarray}
In this notation the NLO pseudoscalar meson masses are \cite{gasser85a}
\begin{eqnarray} 
   M_\pi^2 &=& \chi_\pi \left\{ 
                1 + q_1 \overline{\chi} + q_2 \chi_\pi + \mu_\pi
                  - {1 \over 3} \mu_\eta \right\} 
                                                   \nonumber           \\
   M_K^2   &=& \chi_K \left\{ 
                1 + q_1 \overline{\chi} + q_2 \chi_K 
                  + {2 \over 3} \mu_\eta \right\} 
                                                   \nonumber           \\
   M_\eta^2 &=& \chi_\eta \left\{ 
                 1 + q_1 \overline{\chi} + q_2 \chi_\eta + 2 \mu_K
                   - {4 \over 3} \mu_\eta \right\} 
                                                      \nonumber        \\ 
           & & {}
               + \chi_\pi \left\{ -\mu_\pi + {2 \over 3} \mu_K 
                                          + {1 \over 3} \mu_\eta
                          \right\} 
               + q_3 ( \chi_s - \chi_l)^2 
                                                      \label{Meta} 
\end{eqnarray} 
where 
\begin{eqnarray} 
   q_1 &=& {48 \over f_0^2} (2 L_6 - L_4) 
                                                      \nonumber       \\
   q_2 &=& {16 \over f_0^2} (2 L_8 - L_5) 
                                                      \nonumber       \\
   q_3 &=& {128 \over 9 f_0^2} (3 L_7 + L_8) \,.
\end{eqnarray}


\subsection{Wilson Flow scale, $t_0$}


In \cite{bar13a}, eq.(4.10), B\"ar and Golterman give the form  expected
for the quantity $t_0$ at NNLO in chiral perturbation theory.
\begin{eqnarray} 
   t_0 &=& t_{\rm 0,ch} \left [ 1
           + {k_1 \over (4 \pi f_0)^2} ( 2 M_K^2 + M_\pi^2)  \right. 
                                                      \nonumber       \\
       & & {} 
           + {1 \over (4 \pi f_0)^2 } 
             \left( (3 k_2 - k_1) M^2_\pi \mu_\pi + 4 k_2M^2_K \mu_K
           + {k_1 \over 3} (M^2_\pi - 4 M^2_K) \mu_\eta
           + k_2 M^2_\eta \mu_\eta \right) 
                                                      \nonumber       \\
       & & \left. {} 
           + {k_4 \over (4 \pi f_0)^4} ( 2 M_K^2 + M_\pi^2)^2 
           + {k_5 \over (4 \pi f_0)^4} (M^2_K - M^2_\pi)^2 \right] 
\label{original} 
\end{eqnarray} 
The free parameters in this expression are $k_1, k_2, k_4, k_5$.
Most terms in the expression are obviously symmetric,
but the $k_1$ term has been written in a way that obscures its symmetry. 

Using eqs.~(\ref{Meta}) to translate eq.~(\ref{original})
into $\chi$ variables%
\footnote{In this Appendix we work to order $\chi_q^2$,
dropping terms of order $\chi_q^3$.}  
\begin{eqnarray} 
   t_0 &=& t_{\rm 0,ch} 
            \left [ 1 + {3 k_1 \over (4 \pi f_0)^2}\, \overline{\chi}
            + {k_2 \over (4 \pi f_0)^2 } 
            \left(
              3 \chi_\pi \mu_\pi + 4 \chi_K \mu_K + \chi_\eta \mu_\eta
            \right)
            \right.  
                                                      \nonumber       \\
       & & \left. \hspace*{0.50in}
           + {9 k_4^\prime \over (4 \pi f_0)^4}\, \overline{\chi}^2 
           + {k_5^\prime \over 4 (4 \pi f_0)^4} (\chi_s - \chi_l)^2 
           \right] 
\label{chiform} 
\end{eqnarray} 
with 
\begin{eqnarray} 
   k_4^\prime &=& k_4 + {\textstyle {1 \over 3}} (4 \pi f_0)^2 k_1 (q_1+q_2)
                                                      \nonumber       \\
   k_5^\prime &=& k_5 + {\textstyle {2 \over 3}} (4 \pi f_0)^2 k_1 q_2 \,.
\end{eqnarray} 
The expression in eq.~(\ref{chiform}) is simpler and more explicitly 
symmetric than eq.~(\ref{original}). 
 
As usual, we get a further simplification if we restrict ourselves
to the line of constant $\overline{\chi}$, 
\begin{eqnarray}
   t_0 &=& T \left[ 
           1 + {k_2 \over (4 \pi f_0)^4} 
           \left( 3 \chi_\pi^2 \ln {\chi_\pi \over \overline{\chi}}
                  + 4 \chi_K^2 \ln {\chi_K \over \overline{\chi}}
                  + \chi_\eta^2 \ln {\chi_\eta \over \overline{\chi}} 
           \right)  \right.
                                                      \nonumber       \\
      & & \left. \hspace*{0.375in}
           + {k_5^{\prime\prime} \over 4 (4 \pi f_0)^4} 
                         (\chi_s - \chi_l)^2 
           \right] 
\label{mconst} 
\end{eqnarray}
with 
\begin{eqnarray}
   k_5^{\prime \prime}
      = k_5^\prime + {20 \over 9} k_2 \ln {\overline{\chi} \over \Lambda^2} 
\end{eqnarray}
and 
\begin{eqnarray}
   T = t_{\rm 0,ch} 
       \left[ 1 + {3 k_1 \over (4 \pi f_0)^2}\, \overline{\chi}
              + {8 k_2 \over (4 \pi f_0)^4 } \overline{\chi}^2 
                        \ln {\overline{\chi} \over \Lambda^2} 
              + {9 k_4^\prime \over (4 \pi f_0)^4}\, \overline{\chi}^2
       \right] 
\end{eqnarray}
being the value of $t_0$ on the symmetric line. We can Taylor
expand eq.~(\ref{mconst}) about the symmetric point, the result is
\begin{equation} 
   t_0 = T \left[ 
     1 + {1 \over (4 \pi f_0)^4} { \textstyle ( {5 \over 6} k_2 + {1 \over 4} 
           k_5^{\prime\prime} )} (\chi_s - \chi_l)^2 + \cdots 
           \right]  \,.
 \end{equation} 
 As expected, there is no linear term, and the first term we see 
is quadratic in the $SU(3)$ breaking.


\section{Systematic errors}
\label{systematic}


We follow here the more general discussion given in Appendix~A of
\cite{horsley12a}.


\subsection{Finite lattice volume}

\begin{figure}[htb]
   \vspace*{0.15in}
   \begin{center}
      \includegraphics[width=9.50cm]
      {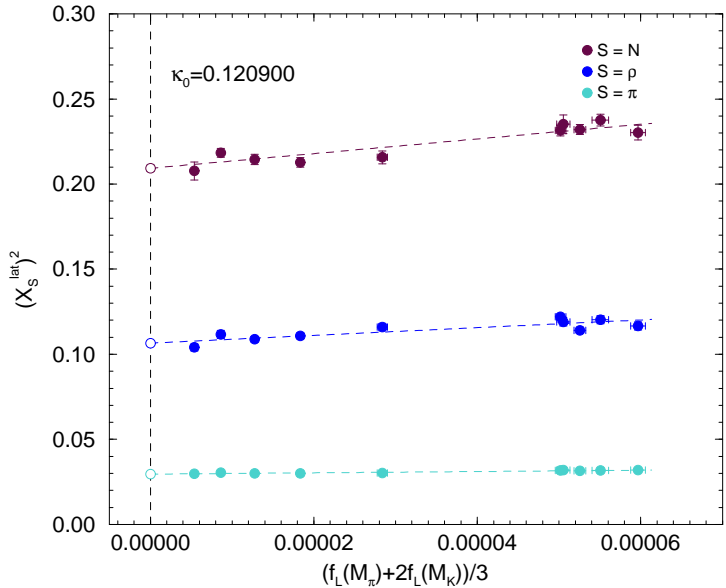}
   \end{center} 
   \caption{$(X_S^{\rm lat})^2$ versus $(f_L(M_\pi) + 2f_L(M_K))/3$ for
            $(\beta, \kappa_0) = (5.50, 0.120900$, with
            $N$ (squares), $\rho$ (diamonds) and $\pi$ (upper triangles).
            The left-most clusters of
            points are from the $32^3 \times 64$ and $48^3\times 96$ 
            lattices, while the right cluster is from $24^3\times 48$ 
            lattices. The dashed lines are linear fits.}
\label{b5p50_ampi2ompiL-3o2exp-mpiL+2amK2omKL-3o2exp-mKLo3_aX2}
\end{figure}

Clearly the argument given in section~\ref{extrap_singlet} that $X_S$ 
is flat along the $SU(3)$ flavour symmetric point holds for any volume.
As discussed in \cite{bietenholz11a} for an estimate of finite volume
effects, a suitable expression is given by
\begin{eqnarray}
   X_S^2(L) 
      = X_S^2 \left( 1 + c_S \textstyle{1\over 3}[ f_L(M_\pi) + 2 f_L(M_K) ]
                \right) \,.
\end{eqnarray}
Lowest order $\chi$PT, \cite{colangelo05a,alikhan03a}
indicates that reasonable functional forms for $f_L(M)$ are
\begin{eqnarray}
   f_L(M) &=& (aM)^2 {e^{-ML} \over (ML)^{3/2}} \,, \qquad \mbox{meson}\,,
                                                             \nonumber \\
   f_L(M) &=& (aM)^2 {e^{-ML} \over (X_NL)} \,, \qquad \mbox{baryon} \,.
\end{eqnarray}
In Fig.~\ref{b5p50_ampi2ompiL-3o2exp-mpiL+2amK2omKL-3o2exp-mKLo3_aX2}
we plot $(f_L(M_\pi) + 2 f_L(M_K))/3$ against $(X_S^{\rm lat})^2$ for 
$S = N$, $\rho$ and $\pi$ on $48^3 \times 96$, $32^3 \times 64$
and additionally $24^3 \times 48$ lattices for $\beta = 5.50$ and
$\kappa_0 = 0.120900$. The fits are linear, with reasonable agreement
to the data. Little finite size effect is seen between the larger
lattice volume results used in the previous analysis and the
extrapolated value here. For the case considered here for 
$(X_\pi, X_N, X_\rho)$ the changes are about $(1.3,2.9,1.5)\%$. 
Taking this as an increase in errors for all the data sets
and performing the same analysis gives the change in the central
values and errors, taken as the systematic error as given
in eq.~(\ref{phys_vals}).


\subsection{$SU(3)$ flavour breaking expansion}


We first note that in
Figs.~\ref{b5p80_mbar_k0p122810_mpsO2o2mpsK2+mps2o3_aX2_48x96},
\ref{b5p50_mbar_k0p120900_mpsO2o2mpsK2+mps2o3_aX2_32x64+48x96}
from the $SU(3)$ flavour symmetric line down to the physical point
lies in the range $|\delta m_l| \lsim 0.01$ (and $|\delta m_s| \lsim 0.02$),
e.g.\ \cite{horsley14a} and that mass `fan plots' (e.g.\ Fig.~5
of \cite{horsley14a}) show little curvature. This is in agreement
with the $SU(3)$ flavour breaking expansion, eq.~(\ref{fit_mpsO})
or (\ref{fit_mNO}). The next order in the expansion is multiplied
by a further $\delta m_q$. So we expect that every increase
in the order leads to a decrease by an order of magnitude or more
(often by a factor $\sim 20$) in the series. So we believe that convergence
is very good for hyperons. (Such an expansion is good compared
to most approaches available to QCD.) 
Nevetheless we have, however, made tests with a linear or quadratic
fit for example for the nucleon in
Fig.~\ref{b5p50_mbar_k0p120900_mpsO2o2mpsK2+mps2o3_aX2_32x64+48x96}
and followed this through the analysis. The final change in central
value for $\sqrt{t_0}$ and $w_0$ was not large, we include it
as a (second) systematic error.


\subsection{Physical scale}


As mentioned in footnote~\ref{emfoot} physical values of hadron
masses have a small electromagetic component. Although we disregard
this in our analysis, we make a small allowance here, and take
$\sqrt{t_0^{\exp}}$, $w_0^{\exp}$ to also have a similar error as $X_\pi$,
i.e.\ a systematic error of $\sim 0.2\%$ due to electromagnetic effects.



\end{document}